\documentclass[12pt]{article}
\usepackage{graphics}
\usepackage{graphicx}
\usepackage{cite}
\begin{document}
\title{Complete Lorentz Transformation\\ of a Charge-Current Density}
\author{Jerrold Franklin\footnote{Internet address:
Jerry.F@TEMPLE.EDU}\\
Department of Physics\\
Temple University, Philadelphia, PA 19122-1801
\date{}}
\maketitle
\begin{abstract}
It is generally assumed in the literature that a Lorentz transformation on a neutral current loop results in a moving current loop with a nonvanishing charge distribution
and an electric dipole moment.
We show in this paper that this is not, in fact, correct. The derivation that leads to the charge distribution was based on an incomplete
Lorentz transformation, which transforms the charge-current four-vector $j^\mu=[\rho({\bf r},t),{\bf j(r},t)]$, but not the space-time four-vector $x^\mu=(t,{\bf r})$.
 We show that completing the Lorentz transformation by using the variable $t'$ in the moving frame, rather than keeping the rest frame time variable $t$, 
results in there being no induced charge density and no resulting electric dipole moment.
\end{abstract}

A large number of papers have been written with the assumption that a moving current loop acquires an induced electric dipole 
moment\cite{tb1,gev,km,vh,a,gh,bt,tb2,v,s,k,sb,gh2,b}
This assumption is generally based on a derivation in Panofsky and Phillips\cite{pp} that purports to demonstrate that a Lorentz transformation on a neutral current loop at rest results in a charged  current loop with an electric dipole moment.  However, the Panofsky and Phillips derivation is wrong because they failed to complete the Lorentz transformation.  Completing the Lorentz transformation by using the the moving frame time variable, rather than keeping the rest frame time variable, results in there being no induced charge density and no resulting electric dipole moment.

A complete Lorentz transformation for a current loop is a two step process. 
The first step  is to Lorentz transform the charge-current density four-vector\footnote {We use units with c=1.} $[\rho({\bf r},t),{\bf j(r},t)]$ 
from the rest system S to a system S$'$ in which the current loop is moving with a velocity $\bf V$.
For a neutral current loop ($\rho=0$) originally at rest, this gives the Lorentz transformation equations
\begin{eqnarray}
\rho'({\bf r},t)&=&\gamma{\bf V\cdot j(r,}t)\label{rp},\\
{\bf j'(r,}t)&=&\gamma{\bf j(r,}t).
\end{eqnarray} 
The positive sign in Eq.~(\ref{rp}) appears because the velocity of the Lorentz transformation 
from S to S$'$ is  $\bf -V$.

This Lorentz transformation seems to have produced a non-vanishing charge distribution in the frame in which the current loop is moving.
On the basis of this, Panofsky and Phillips deduced that a moving current loop would develop an electric dipole moment
due to the charge distribution indicated by Eq.~(\ref{rp}). But the transformation in Eq.~(\ref{rp}) is not a complete Lorentz transformation of the charge-current four-vector. 
The required second step is to Lorentz transform the space-time four-vector $(t,{\bf r})$, so that the transformed $\rho'$ and $\bf j'$ are functions of $t'$ and $\bf r'$.

The charge and current densities, $\rho$ and $\bf j$, are usually idealized as
smooth macroscopic quantities, but the actual physical microscopic charge and current densities in a current loop consist of point conduction electrons moving through a fixed lattice of positively charged ions.  The macroscopic densities are defined by averages  of the microscopic densities over small sampling cells that contain a large number of the moving electrons. 

It is important to use the microscopic densities, and not the simpler macroscopic densities, to understand the details of how the moving electrons form  a current.  Using the macroscopic densities can lead to incorrect and misleading results, which was the case in the Panofsky and Phillips derivation.  

The assumption has usually been made that, because the macroscopic current density is time independent, the second stage of the Lorentz transformation from the rest frame time $t$ to the moving time $t'$ is unnecessary.  But, because the microscopic conduction electrons are moving, completing the Lorentz transformation from $t$ to $t'$ is needed, and changes the final result.

This is similar to the case of the electric field of a moving point charge. There the field of a point charge at rest is time independent, but the Lorentz transformation on the electric field is done in two steps as we described above. However, that procedure was not used by Panofsky and Phillips
for Lorentz transforming a current loop.

To measure the macroscopic charge density $\rho$, we count the number of conduction electrons in a sampling cell, as shown in Fig.~1.  In the loop's rest frame, the velocity $V$ in the figure is zero,
and the electrons are moving with a drift velocity
slowly to the right.
 In order to count the moving electrons correctly in the rest frame, they must all be counted at the same rest frame time. This makes the number of moving electrons the same as the number of fixed ions, and the rest frame charge density is zero.

The rest frame time,
 $t$, and the moving frame time, $t'$, are connected by a Lorentz transformation.
This means that $t'$, the appropriate time for counting electrons in the moving frame, will vary according to the Lorentz transformation equation
\begin{equation}
t'=t/\gamma+Vx',
\label{lt}
\end{equation}
where $x'$ is the distance measured from the back end of the sampling cell.

Since the rest frame time $t$ must remain fixed to keep the wire neutral, the moving frame time $t'$ must vary with position, as seen in Eq.~(\ref{lt}).
This variation of the relevant time variable, $t'$, with position in the sampling cell was the key point missing in the Panofsky and Phillips derivation.

To see the effect of a counting time that varies with distance, we consider the sampling cell  
shown in Fig.~1, consisting of a short stretch of the wire moving with velocity $V$ to the right.
\vspace{-.6in}
\begin{figure}[h]
\hspace{.4in}\includegraphics[width=4.5in]{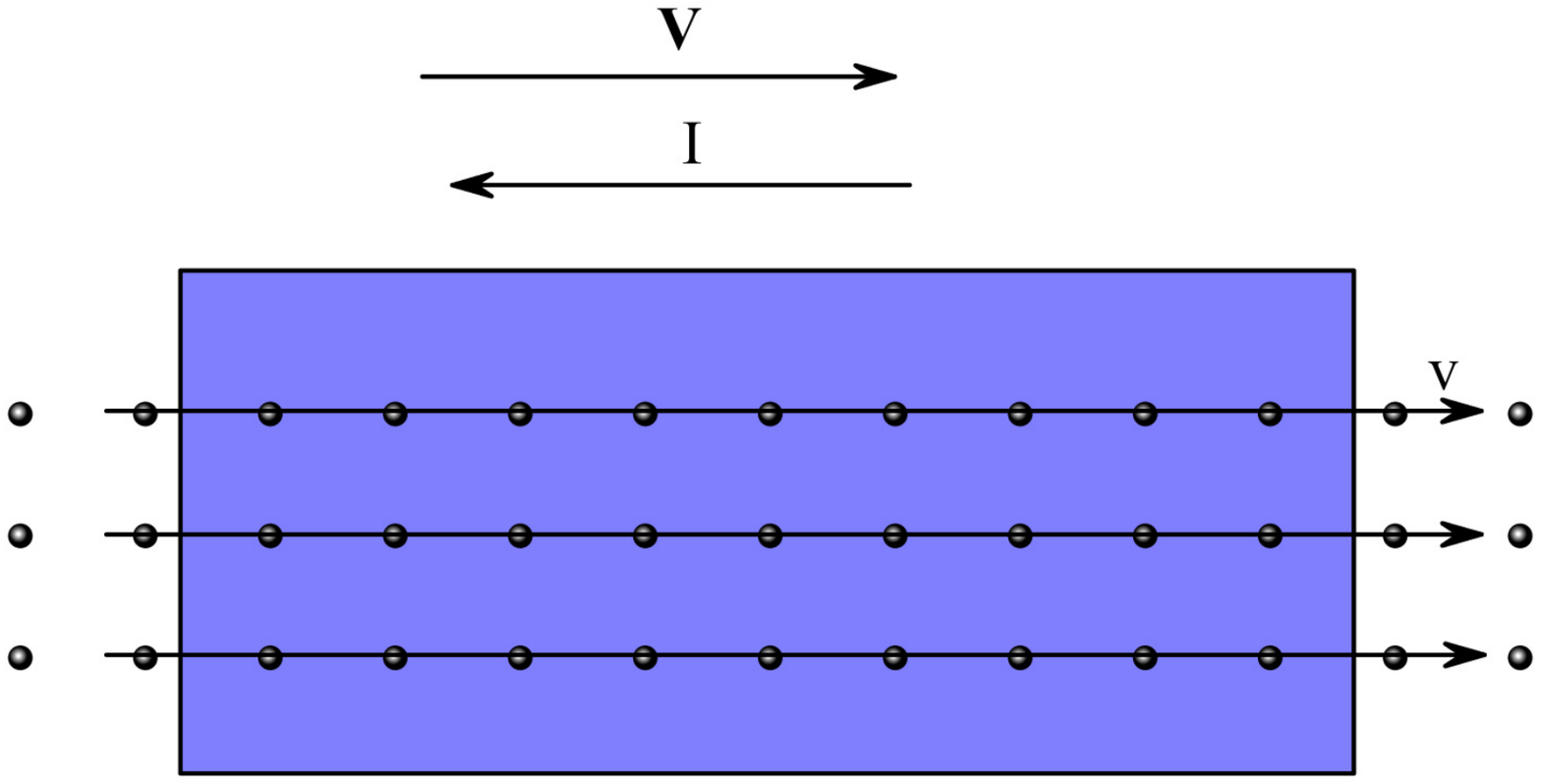} 
\vspace{-.5in}
\caption{Conduction electons moving from left to right in a sampling cell.
The electrons enter the cell at the left end and leave at the right end.}
\end{figure}

In the figure, the electrons are moving from left to right with a small drift velocity  
with respect to the cell, 
corresponding to an electric current flowing from right to left.
In the moving system,
we  should count the electrons at time $t'$,  which, as given by Eq.~(\ref{lt}), increases 
as we count from the left end of the cell to the right.
That means that we would start counting at the left end of the cell and,
by the time we got to the right end, some electrons would have left the cell before we counted them.  We would have counted fewer electrons than were in the cell at any fixed time.
 
The time taken moving from left to right to count the electrons in a sampling cell of length $L'$ 
is given by Eq.~(\ref{lt}) to be 
$\Delta t'=VL'$.
During that time, the 
change in the charge density due to the negatively charged electrons
leaving the right end of the cell would be
\begin{equation}
 \Delta\rho'_2=\frac{j'A'\Delta t'}{A'L'}=\frac{(\gamma jA')(VL')}{A'L'}=\gamma jV,
\end{equation}
where $A'$ is the cross-sectional area and 
$A'L'$ the volume of the sampling cell.
Since negative electrons have left the sampling cell before they were counted, this corresponds to a positive contribution to the charge density.

We have used the subscript 2 in the charge density $\Delta\rho'_2$, because it is the change in the charge density due to the second stage of the Lorentz transformation.
The first stage change in the charge density found by Panofsky and Phillips is given by Eq.~(\ref{rp}) to be
\begin{equation}
 \Delta\rho'_1=-\gamma jV.
\end{equation}
The minus sign arises because $\bf j$ and $\bf V$ are in opposite directions, as seen in Fig.~1.

We see that the net change in charge density, given by Lorentz transforming the space-time variables as well as the charge-current four-vector, is
\begin{equation}
 \Delta\rho'= \Delta\rho'_1+ \Delta\rho'_2=0,
\end{equation}
so a neutral current loop remains neutral when it moves with uniform velocity.
The Panofsky and Phillips derivation found a moving current to be charged because they made the mistake of not completing the Lorentz transformation.

There is another derivation, by Fischer\cite{fi}, that also finds the moving current loop to be charged.  
That derivation does use the moving frame time variable $t'$, but makes the mistake of holding it fixed. However, as we have shown, it is the rest frame time $t$ that must be held fixed so that the current loop will be neutral in its rest frame. Then Eq.~(\ref{lt}) shows that this requires $t'$ to vary with distance, and not be held fixed.

We have shown that a correct Lorentz transformation on a neutral current loop produces a moving, but still neutral, current loop.
This means that the induced charge density found by Panofsky and Phillips, and Fischer is spurious.  With a zero charge density, there would be no induced electric dipole moment in a moving current loop.
This contradicts the large number of papers [1-14] that were generally based on the notion that a moving current loop acquires an electric dipole moment.

One particular consequence is that the claim by Mansuripur\cite{a} that the Lorentz force produces a torque on a moving current loop in the presence of a co-moving point charge was wrong.  It also means that the many conflicting Comments\cite{gh,bt,tb2,v,s,k,sb,gh2} refuting Mansuripur's Physical Review Letter were endeavoring  to resolve a nonexistent problem.  The charged moving current loop that Mansuripur and others tried to reconcile was not the Lorentz transformation of a neutral current loop at rest.

I realize that I disagree with all of my references.  But, actually, those papers just implemented (without question) the Panofsky and Phillips result. None of them have anything like an independent
 derivation of Eq.~(\ref{rp}).  
Thus, although there are fourteen papers using the Panofsky and Phillips result, there is only one derivation, and it is wrong.\footnote{None of the fourteen papers refer back to Fischer's derivation.}

There is another reason why a moving current loop cannot have the charge density given by Eq.~(\ref{rp}). That charge distribution is uniform throughout the conducting wire carrying the current.  But the charge density in a conductor must be only on the surface of the conductor, which is not the case for the charge density in Eq.~(\ref{rp}).
This would be a paradox more compelling than that proposed by Mansuripur\cite{a}.  Fortunately, this paradox is resolved by our demonstration that there is no induced charge density in the moving current loop.

\end{document}